\documentclass[twocolumn,aps]{revtex4}

\usepackage{graphicx}

\begin{document}

\title{First-order phase transitions in superconducting films: A Euclidean model}
\author{C.A. Linhares}
\address{Instituto de F\'{\i}sica, Universidade do Estado do Rio de Janeiro,\\
Rua S\~{a}o Francisco Xavier, 524, 20559-900 Rio de Janeiro, RJ, Brazil}
\author{A.P.C. Malbouisson, Y.W. Milla and I. Roditi}
\address{Centro Brasileiro de Pesquisas F\'{\i}sicas,\\
Rua Dr. Xavier Sigaud, 150, 22290-180 Rio de Janeiro, RJ, Brazil}

\begin{abstract}
In the context of the Ginzburg--Landau theory for critical phenomena, we
consider the Euclidean $\lambda \phi ^4+\eta \phi^6$ model bounded by two
parallel planes, a distance $L$ separating them. This is supposed to
describe a sample of a superconducting material undergoing a first-order
phase transition. We are able to determine the dependence of the transition
temperature $T_{c}$ for the system as a function of $L$. We show that $%
T_{c}(L)$ is a concave function of $L$, in qualitative accordance with some
experimental results. The form of this function is rather different from the
corresponding one for a second-order transition. \\\\\noindent PACS
number(s): 03.70.+k, 11.10.-z
\end{abstract}

\maketitle

\section{Introduction}

In the last few decades, a large amount of work has been done on the
Ginzburg--Landau phenomenological approach to critical phenomena. An account
on the state of the subject and related topics can be found, for instance,
in Refs.~\cite
{halperin,affleck,lawrie1,lawrie2,brezin,radzi,moore,calan,malbouisson,isaque}%
. Questions concerning the existence of phase transitions may also be raised
if one considers the behavior of field theories as a function of spatial
boundaries. The existence of phase transitions would be in this case
associated to some spatial parameters describing the breaking of
translational invariance, for instance, the distance $L$ between planes
bounding the system. Analyses of this type have been recently performed \cite
{malbouisson2,malbouisson3,urucubaca}. In particular, if one considers the
Ginzburg--Landau model confined between two parallel planes, which is
assumed to describe a film of some material, the question of how the
critical temperature depends on the film thickness $L$ can be raised.

Studies on field theory applied to bounded systems have been done in the
literature for a long time. In particular, an analysis of the
renormalization group in finite-size geometries can be found in \cite
{zinn,cardy}. These have been performed to take into account boundary
effects on scaling laws. In another related topic of investigation, there
are systems that present domain walls as defects, created for instance in
the process of crystal growth by some prepared circumstances. At the level
of effective field theories, in many cases, this can be modeled by
considering a Dirac fermionic field whose mass changes sign as it crosses
the defect, meaning that the domain wall plays the role of a critical
boundary separating two different states of the system \cite{fosco1,fosco2}.
Under the assumption that information about general features of the behavior
of systems undergoing phase transitions in absence of external influences
(like magnetic fields) can be obtained in the approximation which neglects
gauge field contributions in the Ginzburg--Landau model, investigations have
been done with an approach different from the renormalization group
analysis. The system confined between two parallel planes has been
considered and using the formalism developed in Refs.~\cite
{malbouisson2,malbouisson3,urucubaca}, the way in which the critical
temperature is affected by the presence of boundaries has been investigated.
In particular, a study has been done on how the critical temperature $T_c$
of a superconducting film depends on its thickness $L$\thinspace \cite
{luciano,luciano1, urucubaca}. In the present paper we perform a further
step, by considering in the same context an extended model, which besides
the quartic field self-interaction, a sextic one is also present. It is well
known that those interactions, taken together, lead to a renormalizable
quantum field theory in three dimensions and which is supposed to describe
first-order phase transitions.

We consider, as in previous publications, that the system is a slab of a
material of thickness $L$, the behavior of which in the critical region is
to be derived from a quantum field theory calculation of the dependence of
the renormalized mass parameter on $L$. We start from the effective
potential, which is related to the renormalized mass through a
renormalization condition. This condition, however, reduces considerably the
number of relevant Feynman diagrams contributing to the mass
renormalization, if one wishes to be restricted to first-order terms in both
coupling constants. In fact, just two diagrams need to be considered in this
approximation: a tadpole graph with the $\phi ^4$ coupling (1 loop) and a
``shoestring'' graph with the $\phi ^6$ coupling (2 loops) (see Fig.1). No
diagram with both couplings occur. The $L$-dependence appears from the
treatment of the loop integrals, as the material is confined between two
planes a distance $L$ apart from one another. We therefore take the space
dimension orthogonal to the planes as finite, the other two being otherwise
infinite. This dimension of finite extent is treated in the momentum space
using the formalism of Ref.~\cite{malbouisson3}.

The paper is organized as follows. In Section II, we present the model and
the description of a bounded system through an adaptation of the Matsubara
formalism suited for our purposes. The contributions from the two relevant
Feynman diagrams to the effective potential are established, as well as an
expression showing the $L$-dependence of the critical temperature. In
Section III, as we wish to compare our theoretical result with some
experimental data, we need first to make a phenomenological evaluation of
the $\phi^6$ coupling constant, based on the analogous derivation made by
Gorkov for the $\phi^4$ constant. The comparison with measurements is
discussed in Section IV. Finally, in Section V we present our conclusions.

\section{The effective potential in the $\phi ^6$ Ginzburg--Landau model}

We start by stating the Ginzburg--Landau Hamiltonian density in a Euclidean $%
D$-dimensional space, now including both $\phi ^4$ and $\phi ^6$
interactions, in the absence of external fields, given by (in natural units, 
$\hbar =c=k_B=1$), 
\begin{equation}
\mathcal{H}=\frac 12\left| \nabla \varphi \right| ^2+\frac 12m_0^2\left|
\varphi \right| ^2-\frac \lambda 4\left| \varphi \right| ^4+\frac \eta
6\left| \varphi \right| ^6,  \label{hamiltonian}
\end{equation}
where we are taking the approximation in which $\lambda >0$ and $\eta >0$
are the \textit{renormalized} quartic and sextic self-coupling constants.
Near criticality, the bare mass is given by $m_0^2=\alpha (T/T_0-1)$, with $%
\alpha >0$ and $T_0$ being a parameter with the dimension of temperature.
Recall that the critical temperature for a first-order transition described
by the hamiltonian above is higher than $T_0$ \cite{lebellac}. This will be
explicitly stated in Eq.~(\ref{tc}) below. We consider the system confined
between two parallel planes, normal to the $x$-axis, a distance $L$ apart
from one another and use Cartesian coordinates $\mathbf{r}=(x,\mathbf{z})$,
where $\mathbf{z}$ is a ($D-1$)-dimensional vector, with corresponding
momenta $\mathbf{k}=(k_x,\mathbf{q}),\mathbf{q}$ being a ($D-1$)-dimensional
vector in momenta space. The generating functional of Schwinger functions is
written in the form 
\begin{equation}
\mathcal{Z}=\int \mathcal{D\varphi }^{*}\mathcal{D\varphi }\exp \left(
-\int_0^Ldx\int d^{D-1}z\,\mathcal{H}\left( \left| \varphi \right| ,\left|
\nabla \varphi \right| \right) \right) ,  \label{partition}
\end{equation}
with the field $\varphi (x,\mathbf{z})$ satisfying the condition of
confinement along the $x$-axis, $\varphi (x\leq 0,\mathbf{z})=\varphi (x\geq
L,\mathbf{z})=$const. Then the field should have a mixed series-integral
Fourier representation of the form 
\begin{equation}
\varphi (x,\mathbf{z})=\sum_{n=-\infty }^\infty c_n\int d^{D-1}q\,b(\mathbf{q%
})e^{-i\omega _nx-i\mathbf{q}\cdot \mathbf{z}}\tilde{\varphi}(\omega _n,%
\mathbf{q}),  \label{fourier}
\end{equation}
where $\omega _n=2\pi n/L$ and the coefficients $c_n$ and $b(\mathbf{q})$
correspond respectively to the Fourier series representation over $x$ and to
the Fourier integral representation over the ($D-1$)-dimensional $\mathbf{z}$%
-space. The above conditions of confinement of the $x$-dependence of the
field to a segment of length $L$ allow us to proceed, with respect to the $x$%
-coordinate, in a manner analogous as is done in the imaginary-time
Matsubara formalism in field theory and, accordingly, the Feynman rules
should be modified following the prescription 
\begin{equation}
\int \frac{dk_x}{2\pi }\rightarrow \frac 1L\sum_{n=-\infty }^\infty ,\qquad
k_x\rightarrow \frac{2n\pi }L\equiv \omega _n.  \label{prescription}
\end{equation}
We emphasize, however, that we are considering an Euclidean field theory in $%
D$ \emph{purely} spatial dimensions, so we are \emph{not} working in the
framework of finite-temperature field theory. Here, the temperature is
introduced in the mass term of the Hamiltonian by means of the usual
Ginzburg--Landau prescription.

To continue, we use some one-loop results described in \cite
{malbouisson2,malbouisson3,ananos}, adapted to our present situation. These
results have been obtained by the concurrent use of dimensional and
zeta-function analytic regularizations, to evaluate formally the integral
over the continuous momenta and the summation over the frequencies $\omega
_n $. We get sums of polar ($L$-independent) terms plus $L$-dependent
analytic corrections. Renormalized quantities are obtained by subtraction of
the divergent (polar) terms appearing in the quantities obtained by
application of the modified Feynman rules and dimensional regularization
formulas. These polar terms are proportional to $\Gamma $-functions having
the dimension $D$ in the argument and correspond to the introduction of
counterterms in the original Hamiltonian density. In order to have a
coherent procedure in any dimension, those subtractions should be performed
even for those values of the dimension $D$ for which no poles are present.
In these cases a finite renormalization is performed.

In principle, the effective potential for systems with spontaneous symmetry
breaking is obtained, following the Coleman--Weinberg analysis \cite{coleman}%
, as an expansion in the number of loops in Feynman diagrams. Accordingly,
to the free propagator and to the no-loop (tree) diagrams for both
couplings, radiative corrections are added, with increasing number of loops.
Thus, at the 1-loop approximation, we get the infinite series of 1-loop
diagrams with all numbers of insertions of the $\phi ^4$ vertex (two
external legs in each vertex), plus the infinite series of 1-loop diagrams
with all numbers of insertions of the $\phi ^6$ vertex (four external legs
in each vertex), plus the infinite series of 1-loop diagrams with all kinds
of mixed numbers of insertions of $\phi ^4$ and $\phi ^6$ vertices.
Analogously, we should include all those types of insertions in diagrams
with 2 loops, etc. However, instead of undertaking this computation, in our
approximation we restrict ourselves to the lowest terms in the loop
expansion. We recall that the gap equation we are seeking is given by the
renormalization condition in which the renormalized squared mass is defined
as the second derivative of the effective potential $U(\varphi _0)$ with
respect to the classical field $\varphi _0$, taken at zero field, 
\begin{equation}
\left. \frac{\partial ^2U(\varphi _0)}{\partial \varphi _0{}^2}\right|
_{\varphi _0=0}=m^2.  \label{renorm}
\end{equation}
Within our approximation, we do not need to take into account the
renormalization conditions for the interaction coupling constants, i.e.,
they may be considered as already renormalized when they are written in the
Hamiltonian. At the 1-loop approximation, the contribution of loops with
only $\phi ^4$ vertices to the effective potential is obtained directly from 
\cite{malbouisson3}, as an adaptation of the Coleman--Weinberg expression
after compactification in one dimension, 
\begin{eqnarray}
U_1(\phi ,L) &=&\mu ^D\sqrt{a}\sum_{s=1}^\infty \frac{(-1)^{s+1}}{2s}%
g_1^s\phi _0^{2s}  \nonumber \\
&&\times \sum_{n=-\infty }^\infty \int \frac{d^{D-1}k}{\left( \mathbf{k}%
^2+an^2+c^2\right) ^s}.  \label{poteffi4}
\end{eqnarray}
In the above formula, in order to deal with dimensionless quantities in the
regularization procedure, we have introduced parameters $c^2=m^2/4\pi ^2\mu
^2$, $a=(L\mu )^{-2}$, $g_1=-\lambda /16\pi ^2\mu ^{4-D}$ and $\phi
_0=\varphi _0/\mu ^{D/2-1}$, where $\varphi _0$ is the normalized vacuum
expectation value of the field (the classical field) and $\mu $ is a mass
scale. The parameter $s$ counts the number of vertices on the loop.

It is easily seen that only the $s=1$ term contributes to the
renormalization condition (\ref{renorm}). It corresponds to the tadpole
diagram. It is then also clear that all $\phi ^6$-vertex and mixed $\phi ^4$%
- and $\phi ^6$-vertex insertions on the 1-loop diagrams do not contribute
when one computes the second derivative of similar expressions with respect
to the field at zero field: only diagrams with two external legs should
survive. This is impossible for a $\phi ^6$-vertex insertion at the 1-loop
approximation, therefore the first contribution from the $\phi ^6$ coupling
must come from a higher-order term in the loop expansion. Two-loop diagrams
with two external legs and only $\phi ^4$ vertices are of second order in
its coupling constant, and we neglect them, as well as all possible diagrams
with vertices of mixed type. However, the 2-loop shoestring diagram, with
only one $\phi ^6$ vertex and two external legs is a first-order (in $\eta $%
) contribution to the effective potential, according to our renormalization
criterion.

\begin{figure}[t]
\includegraphics[{height=4.0cm,width=2.5cm,angle=270}]{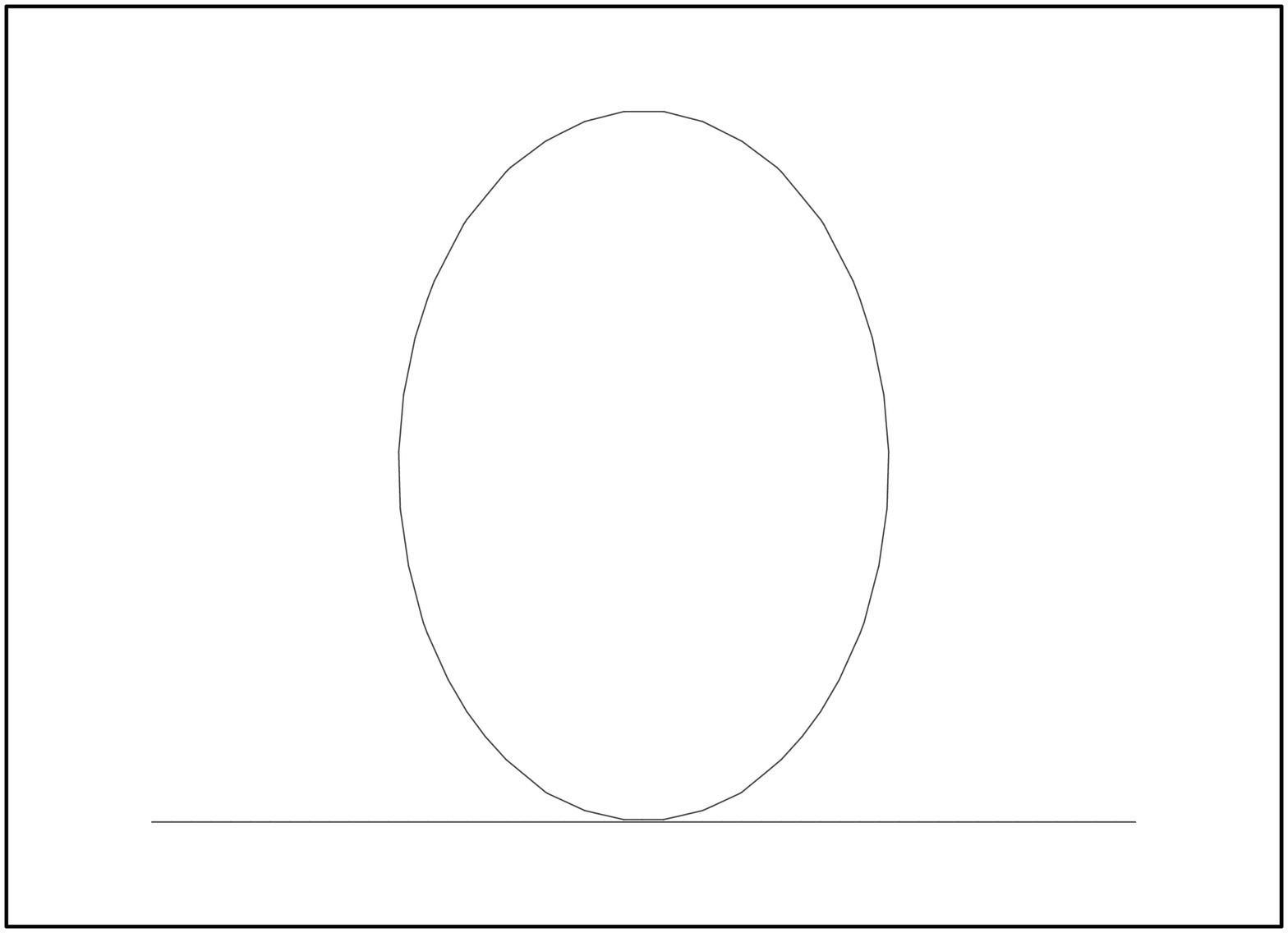} %
\includegraphics[{height=4.0cm,width=2.5cm,angle=270}]{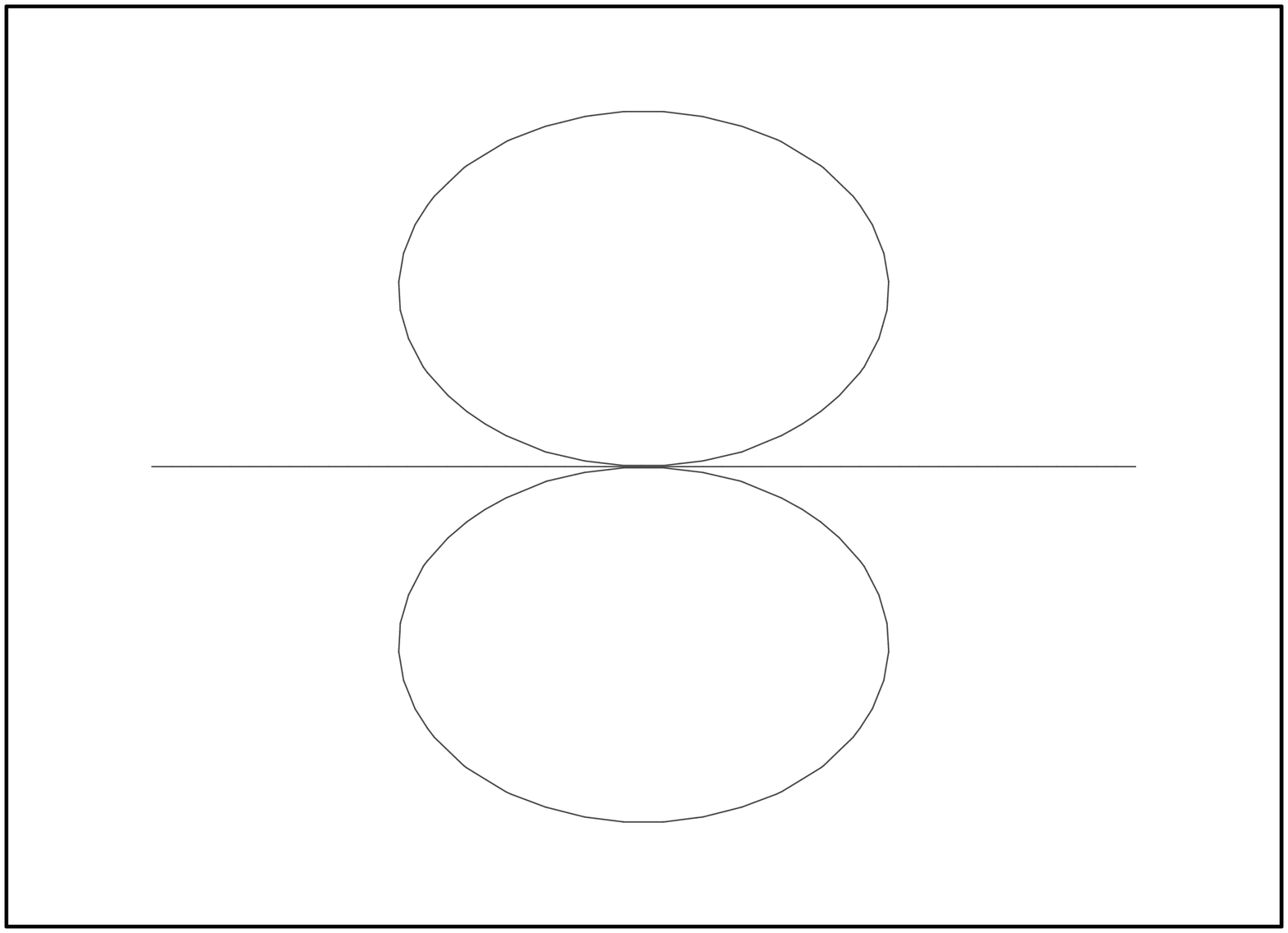}
\caption{Diagrams contributing to the renormalized mass at lowest order in
the coupling constants.}
\end{figure}

Therefore the renormalized mass is defined at first order in both coupling
constants, by the contributions of radiative corrections from only two
diagrams: the tadpole and the shoestring diagrams. The tadpole contribution
reads (putting $s=1$ in Eq. (\ref{poteffi4})), 
\begin{equation}
U_1(\phi _0,L)=\mu ^D\sqrt{a}\frac 12g_1\phi _0^2\sum_{n=-\infty }^\infty
\int \frac{d^{D-1}k}{\mathbf{k}^2+an^2+c^2}.  \label{tadpole}
\end{equation}
The integral on the $D-1$ non-compactified momentum variables is performed
using the dimensional regularization formula 
\begin{equation}
\int \frac{d^dk}{k^2+M}=\frac{\Gamma \left( 1-\frac d2\right) \pi ^{d/2}}{%
M^{1-d/2}};  \label{regdim}
\end{equation}
for $d=D-1$, we obtain 
\begin{eqnarray}
U_1(\phi _0,L)&=&\mu ^D\sqrt{a}\frac{\pi ^{(D-1)/2}}2g_1\phi _0^2\Gamma
\left( \frac{3-D}2\right)  \nonumber \\
& &\times\sum_{n=-\infty }^\infty \frac 1{(an^2+c^2)^{(3-D)/2}}.
\end{eqnarray}
The sum in the above expression may be recognized as one of the
Epstein--Hurwitz zeta-functions, $Z_1^{c^2}(\frac{3-D}2;a)$, which may be
analytically continued to \cite{elizalde} 
\begin{eqnarray}  \label{epstein}
Z_1^{c^2}(\nu ;a)&=&\frac{2^{\frac{2\nu +1}{2}}\pi ^{\frac{4\nu -1}{2}}}{%
\sqrt{a}\Gamma (\nu )}\left[ 2^{\nu -3/2}\left( \frac m\mu \right) ^{1-2\nu
}\Gamma \left( \nu -\frac 12\right) \right.  \nonumber \\
& & \left. +2\sum_{n=1}^\infty \left( \frac m{\mu ^2Ln}\right) ^{1/2-\nu
}K_{\nu -1/2}(mnL)\right] ,  \nonumber \\
\end{eqnarray}
where the $K_\nu $ are Bessel functions of the third kind. The tadpole part
of the effective potential is then 
\begin{eqnarray}  \label{tadeff}
U_1(\phi _0,L) &=&\frac{\mu ^Dg_1\phi _0^2}{\left( 2\pi \right) ^{D/2-2}}%
\left[ 2^{-\frac{D+1}{2}}\left( \frac m\mu \right) ^{D-2}\Gamma \left(
1-\frac D2\right) \right.  \nonumber \\
&&\left. +\sum_{n=1}^\infty \left( \frac m{\mu ^2Ln}\right)
^{D/2-1}K_{D/2-1}(mnL)\right] .  \nonumber \\
\end{eqnarray}

We now turn to the 2-loop shoestring diagram contribution to the effective
potential, using again the Feynman rule prescription for the compactified
dimension. It reads 
\begin{eqnarray}
U_2(\phi ,L) &=&\mu ^{2D-2}a\frac 12g_2\phi _0^2  \nonumber  \label{U2} \\
&&\times \left[ \Gamma \left( \frac{3-D}2\right) \pi
^{(D-1)/2}Z_1^{c^2}\left( \frac{3-D}2;a\right) \right] ^2,  \nonumber \\
&&
\end{eqnarray}
where $g_2=\eta /128\pi ^4\mu ^{6-2D}$. After subtraction of the polar term
coming from the first term of Eq. (\ref{epstein}) we get 
\begin{eqnarray}
U_1^{\mathrm{(Ren)}}(\varphi _0,L) &=&\frac{-\lambda \varphi _0^2}{4(2\pi
)^{D/2}}\sum_{n=1}^\infty \left( \frac m{\mu ^2nL}\right)
^{D/2-1}K_{D/2-1}(mnL)  \nonumber  \label{tadeff1} \\
&&
\end{eqnarray}
and 
\begin{eqnarray}
U_2^{\mathrm{(Ren)}}(\varphi _0,L) &=&\frac{\eta \varphi _0^2}{4(2\pi )^D}%
\left[ \sum_{n=1}^\infty \left( \frac m{nL}\right)
^{D/2-1}K_{D/2-1}(mnL)\right] ^2.  \nonumber  \label{U2Ren} \\
&&
\end{eqnarray}
Thus the full renormalized effective potential is given by 
\begin{equation}
U(\varphi _0,L)=\frac 12m_0^2\varphi _0^2-\frac \lambda 4\varphi _0^4+\frac
\eta 6\varphi _0^6+U_1^{\text{\textrm{(Ren)}}}+U_2^{\text{\textrm{(Ren)}}}.
\label{URen}
\end{equation}

The renormalized mass with both contributions then satisfies an $L$%
-dependent generalized Dyson--Schwinger equation, 
\begin{eqnarray}
m^2(L) &=&m_0^2-\frac \lambda {2\left( 2\pi \right) ^{D/2}}\sum_{n=1}^\infty
\left( \frac m{nL}\right) ^{D/2-1}K_{D/2-1}(mnL)  \nonumber  \label{massren1}
\\
&&+\frac \eta {4(2\pi )^D}  \nonumber \\
&&\times \left[ \sum_{n=1}^\infty \left( \frac m{nL}\right)
^{D/2-1}K_{D/2-1}(mnL)\right] ^2.  \nonumber \\
&&
\end{eqnarray}
Thus, the effective potential (\ref{URen}) is rewritten in the form 
\begin{equation}
U(\varphi _0)=\frac 12m^2(L)\varphi _0^2-\frac \lambda 4\varphi _0^4+\frac
\eta 6\varphi _0^6,  \label{potencial}
\end{equation}
where it is assumed that $\lambda ,\eta >0$, a necessary condition for the
existence of a first-order phase transition associated to the potential (\ref
{potencial}). Then, a first-order transition occurs when all the three
minima of the potential are simultaneously on the line $U(\varphi _0)=0$.
This gives the condition 
\begin{equation}
m^2(L)=\frac{3\lambda ^2}{16\eta }.  \label{condicao}
\end{equation}

Notice that the value $m=0$ is excluded in the above condition, for it
corresponds to a second-order transition. For $D=3$, which is the physically
interesting situation of the system confined between two parallel planes
embedded in a 3-dimensional Euclidean space, the Bessel functions entering
in the above equations have an explicit form, $K_{1/2}(z)=\sqrt{\pi }e^{-z}/%
\sqrt{2z}$, which replaced in Eq.(\ref{massren1}), performing the resulting
sum, and reminding that $m_0^2=\alpha (T/T_0-1)$, gives 
\begin{eqnarray}
m^2(L) &=&\alpha \left( \frac T{T_0}-1\right) +\frac \lambda {8\pi }\frac
1L\ln (1-e^{-m(L)L})  \nonumber  \label{massren2} \\
&&+\frac{\eta \pi }{8(2\pi )^3L^2}\left[ \ln (1-e^{-m(L)L})\right] ^2. 
\nonumber \\
&&  \label{massren2}
\end{eqnarray}
In Eq.(\ref{condicao}) $m(L)$ may have any strictly positive value and this
condition ensures that we are on a point on the critical line for a
first-order phase transition. Then introducing the value of the mass, Eq.(%
\ref{condicao}), in Eq.(\ref{massren2}), we obtain the critical temperature 
\begin{widetext} 
\begin{equation}
T_c(L) =T_c\left\{1-\left(1+\frac{3\lambda ^2}{16\eta \alpha}\right)^{-1}
\left[\frac \lambda {8\pi
\alpha L}\ln (1-e^{-\sqrt{\frac{3\lambda ^2}{16\eta }}L})
 +\frac \eta {64\pi ^2\alpha L^2}\left( \ln (1-e^{-\sqrt{\frac{
3\lambda ^2}{16\eta }}L})\right) ^2\right]\right\} ,  \label{massren3}
\end{equation}
\end{widetext}
where 
\begin{equation}
T_c=T_0\left( 1+\frac{3\lambda }{16\eta \alpha }\right)   \label{tc}
\end{equation}
is the bulk ($L\rightarrow \infty $) critical temperature for the
first-order phase transition.

\section{Phenomenological evaluation of the constant $\eta$}

Our aim in this section is to generalize Gorkov's \cite
{gorkov,abrikosov,kleinert} microscopic derivation done for the $\lambda
\varphi ^4$ model in order to include the additional interaction term $\eta
\varphi ^6$ in the free energy. Here, our interest is to determine the
phenomenological constant $\eta $ as a function of the microscopic
parameters of the material, in an analogous way as it has been done for the
constant $\lambda $ in the $\lambda \varphi ^4$ model. Using Gorkov's
equations combined with the self-consistent gap condition \cite{abrikosov}
the free energy density may be written in terms of the gap energy $\Delta (x)
$ as 
\begin{eqnarray}
f(\Delta ) &=&N(0)\left[ \xi _0^2|\nabla \Delta |^2+\left( \frac
T{T_0}-1\right) |\Delta |^2\right.   \nonumber  \label{G1} \\
&&\left. +\frac{3\xi _0^2}{\hbar ^2v_F^2}|\Delta |^4+\frac{1674\,\zeta
(5)\,\xi _0^4}{147\,\hbar ^4v_F^4\,\zeta ^2(3)}|\Delta |^6\right] , 
\nonumber \\
&&
\end{eqnarray}
where $N(0)$ is the density of states at the Fermi surface, $\xi _0$ is the
coherence length, $v_F$ the Fermi velocity, and $\zeta (x)$ is the Riemann
zeta-function. $N(0)$ and $\xi _0$ are given by 
\begin{equation}
N(0)=\frac 1{4\pi ^2k_BT_F}\left( \frac{p_F}\hbar \right) ^3\;,\qquad \xi
_0\approx 0.13\frac{\hbar v_F}{k_BT_0},  \label{parametros}
\end{equation}
where $T_F$ is the Fermi temperature and $k_B$ is Boltzmann's constant. $T_0$
is the temperature parameter introduced in Eq.~(\ref{hamiltonian}) that can
be obtained from the first-order bulk critical temperature by means of Eq.~(%
\ref{tc}). Introducing the order parameter $\varphi =\sqrt{2N(0)}\xi
_0\Delta $ in Eq. (\ref{G1}) we obtain 
\begin{widetext}
\begin{equation}
f(\varphi) =\frac 12|\nabla \varphi |^2+\frac 1{2\xi _0^2}\left( \frac
T{T_0}-1\right) |\varphi | ^2+\frac 3{4\,\hbar ^2v_F^2\,\xi _0^2\,N(0)|}\varphi |^4+
\frac 16\frac{1674\,\zeta (5)}{196\,N(0)^2\,\hbar ^4v_F^4\,\zeta ^2(3)\,\xi
_0^2}|\varphi |^6.
\label{G2}
\end{equation}
\end{widetext}

In order to be able to compare our results with some experimental
observation, we should restore SI units (remember that so far we have used
natural units, $c=\hbar =k_B=1$) \cite{kleinert}. In SI units, the exponent
in the partition function (\ref{partition}) has a factor $1/k_BT_0$. Then,
we must divide by $k_BT$ the free energy density in Eq.~(\ref{G2}).
Moreover, we rescale the fields and coordinates by $\varphi _{\text{new}}=%
\sqrt{\xi _0/k_BT_0}\varphi $ and $x_{\text{new}}=x/\xi _0$, which gives the
dimensionless energy density and, comparing with Eq.~({\ref{hamiltonian}),
we can identify the phenomenological dimensionless constants $\lambda $, $%
\eta $, and $m_0$, with $\alpha =1$ \cite{kleinert}, 
\begin{equation}
\lambda \approx 111.08\,\left( \frac{T_0}{T_F}\right) ^2,\;\;\;\;\eta
\approx 0.04257\,\lambda ^2,\;\;\;\;m_0^2=\frac T{T_0}-1.
\label{parametros3}
\end{equation}
}

By replacing the above constants in Eq.~(\ref{massren3}), we get the
critical temperature as a function of the film thickness and in terms of
microscopic tabulated parameters for specific materials.

\section{Comparison with experimental data}

We remark that Gorkov's original derivation of the phenomenological
constants is valid only for perfect crystals, where the electron mean free
path $l$ is infinite. However, we know that in many superconductors the
attractive interaction between electrons (necessary for pairing) is brought
about indirectly by the interaction between the electrons and the vibrating
crystal lattice (the phonons). Considering that this interaction will be
greater if we have impurities within the crystal lattice, consequently the
electron mean free path is actually finite. The Ginzburg--Landau
phenomenological constants $\lambda $ and $\eta $ and the coherence length
are somehow related to the interaction of the electron pairs with the
crystal lattice and the impurities. A way of taking these facts into account
preserving the form of the Ginzburg--Landau free energy is to modify the
intrinsic coherence length and the coupling constants. Accordingly \cite
{kleinert}, $\xi _0\rightarrow r^{1/2}\xi _0$, $\lambda \rightarrow
2r^{-3/2}\lambda $ and $\eta \rightarrow 4r^{-3}\eta $, where $r\approx
0.18R^{-1}$, with $R=\xi _0/l$. Then, Eq. (\ref{massren3}) becomes 
\begin{widetext}
\begin{equation}
T_c(L) =T_{c}\left\{1-\left( 1+ \frac{3 \lambda ^2}{16\eta}\right)^{-1}
\left[ \frac{2R\lambda }{0.18 \cdot 8\pi }
\frac{\xi _0}L\ln (1-e^{-\frac L{\xi _0}\sqrt{\frac{3\lambda ^2}{16\eta }
\frac R{0.18}}})
 +\frac{4R^2\eta }{0.18^2 32\pi ^2}\left( \frac{\xi _0}L\right)
^2\left( \ln (1-e^{-\frac L{\xi _0}\sqrt{\frac{3\lambda ^2}{16\eta }\frac
R{0.18}}})\right) ^2\right]\right\}.
\label{Tcritical2}
\end{equation}
\end{widetext}

We consider that other effects, such that of the substrate over which the
superconductor film is deposited, should be taken into account. In the
context of our model, however, we are not able to describe such effects at a
microscopic level. We therefore assume that they will be translated in
changes on the values of the coupling constants $\lambda $ and $\eta $. So,
we propose as an \emph{Ansatz} the rescaling of the constants in the form $%
\lambda \rightarrow a\lambda $ and $\eta \rightarrow a^2\eta $. We may still
combine both parameters $R$ and $a$ as $r=aR$. Eq.~(\ref{Tcritical2}) is
then written as 
\begin{widetext}
\begin{equation}
T_c(L) =T_c\left\{1-\left( 1+ \frac{3 \lambda ^2}{16\eta}\right)^{-1}
\left[ \frac{2r\lambda }{0.18 \cdot 8\pi }
\frac{\xi _0}L\ln (1-e^{-\frac L{\xi _0}\sqrt{\frac{3\lambda ^2}{16\eta }
\frac R{0.18}}})
 +\frac{4r^2\eta }{0.18^2 32\pi ^2}\left( \frac{\xi _0}L\right)
^2\left( \ln (1-e^{-\frac L{\xi _0}\sqrt{\frac{3\lambda ^2}{16\eta }\frac
R{0.18}}})\right) ^2\right]\right\}.
\label{Tcritical3}
\end{equation}
\end{widetext}

In Fig. 2 we plot Eq. (\ref{Tcritical3}) to show the behavior of the
transition temperature as a function of the thickness for a film made from
aluminum. The values for Al of the Fermi temperature and the bulk critical
temperature are $T_F=13.53\times 10^4$ K and $T_c=1.2$ K, respectively. 
\begin{figure}[t]
\includegraphics[{height=7.0cm,width=8.5cm,angle=360}]{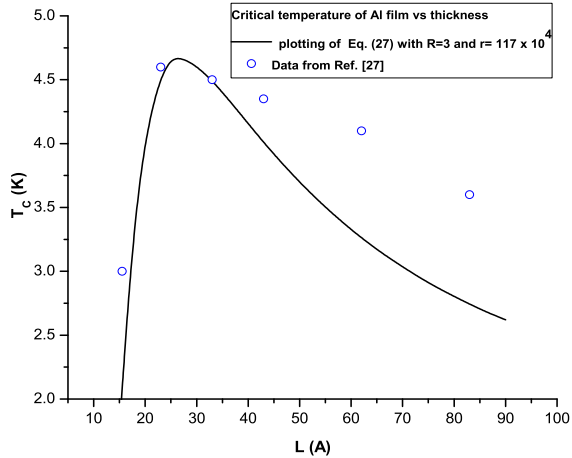}
\caption{Critical temperature $T_c$(K) as function of thickness $L$ ($\AA $%
), from Eq.(\ref{Tcritical3}) and data from Ref.~\protect\cite{strongin} for
a superconducting film made from aluminum.}
\end{figure}
We see from the figure that the critical temperature grows from zero at a
nonnull minimal allowed film thickness above the bulk transition temperature 
$T_c$ as the thickness is enlarged, reaching a maximum and afterwards
starting to decrease, going asymptotically to $T_c$ as $L\rightarrow \infty $%
. We also plot for comparison some experimental data obtained from Ref.~\cite
{strongin}. We see that our theoretical curve is in qualitatively good
agreement with measurements, especially for thin films.

The experimental evidence showing that in some superconducting films the
transition temperature is well above the bulk one has been reported in the
literature since the 1950s and 60s \cite
{buckel,strongin2,strongin5,strongin6}. On the theoretical side, a formula
for the transition temperature was written within BCS theory in terms of the
electron-phonon coupling constant, the Debye temperature and the Coulomb
coupling constant \cite{mcmillan}. This formula was used to explain observed
increases in the critical temperature of thin composite films consisting of
alternating layers of dissimilar metals \cite{strongin2}. In Ref.~\cite
{dickey} a molecular-dynamic technique was applied to obtain the phonon
frequency spectrum which led to the same results. Mechanisms accounting for
the sharp drop in $T_c(L)$ for very thin films were also discussed in Ref.~%
\cite{strongin}. The authors conclude that the most important influence on $%
T_c(L)$ was the interaction of the film with the substrate, described by a
model in Ref.~\cite{cooper}.

It is interesting that in recent reports on copper oxide
high-transition-temperature superconductors, the critical temperature
depends on the number of layers of CuO$_2$ in a similar way as above: first
it rises with the number of layers and, after reaching a maximum value, then
declines. See \cite{ramallo} and references therein.

This behavior may be contrasted with the one shown by the critical
temperature for a \emph{second-order} transition. In this case, the critical
temperature increases monotonically from zero, again corresponding to a
finite minimal film thickness, going asymptotically to the bulk transition
temperature as $L\rightarrow \infty $. This is illustrated in Fig. 3,
adapted from Ref.~\cite{luciano2}, with experimental data from \cite{itoh}.
(Such behavior has also been experimentally found by some other groups for a
variety of transition-metal materials, see Refs.~\cite
{raffy,minhaj,pogrebnyakov}.) Since in the present work a first-order
transition is explicitly assumed, it is tempting to infer that the
transition described in the experiments of Ref.~\cite{strongin} is first
order. In other words, one could say that an experimentally observed
behavior of the critical temperature as a function of the film thickness may
serve as a possible criterion to decide about the order of the
superconductivity transition: a monotonically increasing critical
temperature as $L$ grows would indicate that the system undergoes a
second-order transition, whereas if the critical temperature presents a
maximum for a value of $L$ larger than the minimal allowed one, this would
be signalling the occurrence of a first-order transition. 
\begin{figure}[t]
\includegraphics[{height=7.0cm,width=8.5cm,angle=360}]{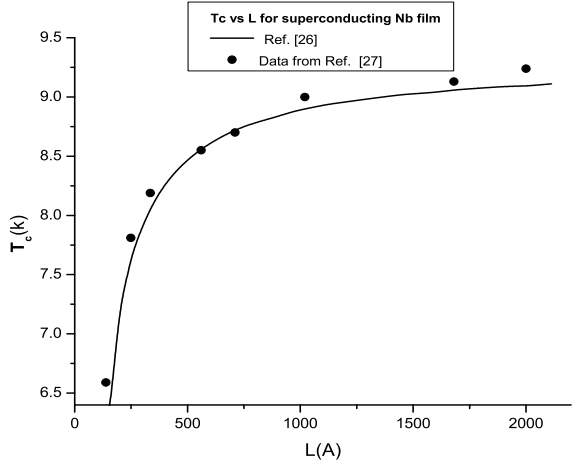}
\caption{Critical temperature $T_c$ (K) as a function of the thickness $L$ ($%
\AA $) for a second-order transition, as theoretically predicted in Ref. 
\protect\cite{luciano2}. Dots are experimental data taken from Ref. 
\protect\cite{itoh} for a superconducting film made from niobium.}
\end{figure}

\section{Conclusions}

As seen in previous works, a superconducting system confined in some region
of space may lose its characteristics if the dimensions of this region
become sufficiently small. This is due to the fact that the critical
temperature depends on these dimensions in such a way that it vanishes below
some finite minimal size. This has been verified in a field-theoretical
framework for a Ginzburg--Landau model describing a second-order phase
transition. In the present paper, we have studied the critical temperature
behavior of a sample of superconducting material in the form of a film, but
we have included in the model a $\phi^6$ self-interaction term, thus
implying that we are now dealing with a first-order transition. In the case
we have treated, a sharply contrasting behavior of the critical temperature,
as a function of the film thickness, was obtained with respect to the
corresponding one for a second-order transition. This possibly indicates a
way of discerning the order of a superconducting transition from
experimental data, according to the profile of the curve $T_c$ vs $L$.

Also importantly, for our derivation of the first-order transition critical
temperature curve, we needed to phenomenologically evaluate the $\phi^6$
coupling constant, which, as far as we know, is not present in the
literature.

Finally, we also remark that in $D=3$, for second-order transitions, one
considers $m=0$ and that leads to the need of a pole-subtraction procedure
for the mass \cite{isaque}. In our case such a procedure is not necessary,
as a first-order transition must occur for a non-zero value of the mass.
This fact, together with the closed formula for the Bessel function for $D=3$%
, allows us to obtain the exact expression (\ref{massren3}) for the critical
temperature.

%%%%%%%%%%%%

\acknowledgments
This work has received partial financial support from CNPq and Pronex.

\end{document}